\documentclass[a4paper]{article} 
\usepackage{amsmath,amsfonts,amsthm}

\theoremstyle{definition}
\newtheorem{definition}{Definition}

\usepackage{bbm}
\numberwithin{equation}{section}

\title{Helicity -- from Clifford to Graphene}
\author{
C. G. B\"ohmer\footnote{c.boehmer@ucl.ac.uk}~ 
and~L. Corpe\footnote{l.corpe@ucl.ac.uk}\\
Department of Mathematics and Institute of Origins\\
University College London, Gower Street, London, WC1E 6BT, UK
}
\date{\today}

\begin{document}
\maketitle
\begin{abstract}
We investigate two seemingly disjoint definitions of helicity, one commonly used in particle physics, the other one used when studying bilinear covariants of Clifford algebras. We can prove that the `mathematical' definition of helicity implies its `physical' counterpart. As an unexpected application of our result we show that the Hamiltonian describing the one-layer superconductor Graphene is proportional to the trace of an operator that is used in the `mathematical' definition of helicity.
\end{abstract}


\section{Introduction} 

The concept of helicity is important in modern physics, it often plays a major role in particle physics, and appears in many other areas of physics: from high energy physics, where it is directly linked to parity violation in the weak interaction, spinorial fields, to solid state physics, where it is used to express the Hamiltonian of the one-layer superconductor Graphene. However, the formalisms used to describe these different physical phenomena can vary greatly from field to field, and as a consequence of this, helicity has several different definitions depending on the area in which it is studied. Although these different versions of the same concept are often intuitively similar, they can take different forms depending on the subject one is specialised in.  In this paper, we will explore the link between two seemingly disjoint ways of thinking of the concept of helicity. Since they are both traditionally used in conjunction with Dirac spinors, we will limit our study to relativistic spin-1/2 particles. In Section 2.1 we will delve into the definition of helicity most commonly used, that of the helicity operator and its corresponding eigenvalues and eigenvectors, which is prevalent in quantum mechanics and related theories. In Section 2.2, we will look into the definition of helicity in terms of a relation between bilinear covariants, which looks at the concept from a more mathematically-oriented point of view, namely through the study of Clifford algebras. In Section 3, which shows our main result, we will establish the link between these disjoint definitions, proving eventually that one definition of helicity implies the other. In the final section, we apply our results to Graphene and outline possible applications.

\section{Preliminaries}

\subsection{Helicity in Physics}

\subsubsection{The helicity operator}

In this section, we will explore the most common version of helicity, as used in quantum theory and related fields. This definition is part of a family of definitions, all of which have a very similar form, but differ in trivial ways, most commonly by factors of 1/2 and $\hbar$, Planck's reduced constant. We will use a definition whereby helicity of a particle takes the values $+1$ or $-1$. 

\begin{definition}{The helicity operator in physics~\cite[p.~42]{Ryder}:}
We define the helicity operator as the mapping $(\vec{\sigma} \cdot \hat{p}) : \mathbb{C} ^2 \rightarrow \mathbb{C}^2 $. It has eigenvalues $+1$ and $-1$ with corresponding eigenvectors $|\phi_R \rangle$ and $|\phi_L \rangle$, the right-handed (positive helicity) and left-handed (negative helicity) eigenstates, respectively.
\label{def1}
\end{definition}

An important point here is that the helicity operator is defined as an inner product between spin projection operators and an operator giving the direction of motion. Namely, we use the Pauli spin matrices  
\begin{align}
  \sigma_1=\begin{pmatrix}0&1\\1&0\end{pmatrix},\quad 
  \sigma_2=\begin{pmatrix}0& -i \\ i & 0 \end{pmatrix},\quad
  \sigma_3=\begin{pmatrix} 1 & 0 \\ 0 & -1 \end{pmatrix},
\end{align} 
and unit momentum $\hat{p} = p/|p|$. The eigenvalue equation for helicity is thus
\begin{align} 
  (\vec{\sigma} \cdot \hat{p}) |\phi_{R/L} \rangle = \pm | \phi_{R/L} \rangle,
\end{align}
where $(\vec{\sigma} \cdot \hat{p}) := \sigma_1 p_1 + \sigma_2 p_2 + \sigma_3 p_2$, where $p_1$, $p_2$ and $p_3$ are the components of $\hat{p}$. We now have an equation which defines the helicity operator, as well as its eigenvectors and eigenvalues.

\subsubsection{Momentum space representation}

We would like to have an explicit form for the eigenvectors $|\phi_L \rangle $ and $ | \phi_R \rangle $ in terms of the components of $\hat{p}$. This unit vector contains no information about the actual momentum of the particle, only information about its direction, the \emph{magnitude} of momentum is of no importance at this stage. Thus, we can choose to write $\hat{p} $ as a general spatial unit vector without loss of generality. It turns out to be convenient to write $\hat{p}$ as a unit vector in spherical polar coordinates 
\begin{align} 
  \hat{p} = \begin{pmatrix} 
    \sin\theta \cos\varphi \\ 
    \sin \theta \sin \varphi \\ 
    \cos \theta \end{pmatrix}, 
  \quad  \theta \in [0, \pi], \quad \varphi \in [0, 2\pi). 
\end{align}
When we substitute this into the definition of the helicity operator given above, we can work out its explicit form 
\begin{align} 
  (\vec{\sigma}\cdot\hat{p}) = 
  \begin{pmatrix} 
    \cos\theta  & \sin\theta e^{-i\varphi} \\ 
    \sin\theta e^{i\varphi} & -\cos\theta 
  \end{pmatrix}. 
  \label{helicity}
\end{align}
It is then straightforward to find the explicit forms for our 2-entry spin eigenvectors in terms of the coordinates of the unit momentum vector. As such, our spinors are now `functions of momentum' and so we call this representation the \emph{momentum space representation}:
\begin{align} 
  |\phi_R \rangle &= 
  \begin{pmatrix} 
    \cos (\theta/2) e^{-i\varphi /2} \\  
    \sin (\theta /2)  e^{i \varphi /2 } 
  \end{pmatrix},
  \label{psiR}\\  
  | \phi_L \rangle &=  
  \begin{pmatrix} 
    \sin (\theta/2) e^{-i\varphi' /2} \\
    - \cos  (\theta /2)  e^{i \varphi' /2 } 
  \end{pmatrix},
  \label{psiL}
\end{align}
where $\varphi' = \varphi + \Delta\varphi$, and $\Delta \varphi$ is an arbitrary phase difference between the left-handed and right-handed solutions.

\subsubsection{Gamma matrices and Dirac spinors}

Spin-1/2 particles (Fermions) are modelled by the \emph{Dirac equation} $-i \gamma^a\partial_a\psi + m\psi = 0$, where $\gamma^a$ are the Dirac gamma matrices which are based on the Pauli matrices defined above. The positive-energy solutions describe `normal' particles while the negative-energy solutions are interpreted as \emph{anti-particles}, both with positive probabilities. The method used to derive this is well summarised in Lounesto~\cite[p.~135]{Lounesto} and every book on Quantum Field Theory.
 
The gamma matrices are important in this area of physics, so we briefly review them here. They are $4\times4$ matrices and represent the mathematical structure of the Clifford algebra $Cl_{1,3}$. They obey the relation $\{\gamma^a,\gamma^b\} := \gamma^a\gamma^b\ + \gamma^b\gamma^a = 2 \eta^{ab} \mathbbm{1}_{4}$, where $\mathbbm{1}_{4}$ stands for the $4\times4$ identity matrix. The curly bracket $\{\ ,\,\}$ is called the anti-commutator, and $\eta^{ab} $ is the Minkowski space-time metric with signature $(+,-,-,-)$. We will use the Weyl basis for the gamma matrices, which we can express as follows in terms of the $2\times2$ identity matrix $\mathbbm{1}_{2}$, the $2\times2$ zero matrix $\mathbb{O}_{2}$ and the Pauli matrices. We can view the gamma matrices as $2\times2$ matrices with $2\times2$ matrix entries
\begin{align} 
  \gamma^0 = 
  \begin{pmatrix} \mathbb{O}_2 & \mathbbm{1}_2 \\ 
    \mathbbm{1}_2 & \mathbb{O}_2 
  \end{pmatrix}, \quad 
  \gamma^a = 
  \begin{pmatrix} \mathbb{O}_2 & \sigma_a \\ 
    - \sigma_ a & \mathbb{O}_2 
  \end{pmatrix}, \quad 
  a = 1,2,3. 
\end{align}
An additional gamma matrix, known as $\gamma^5$ is defined as $ \gamma^5 = i \gamma^0 \gamma^1 \gamma^2 \gamma^3$. The eigenvectors of the Dirac Hamiltonian are known as Dirac spinors. A convenient notation here is the Dirac notation, where $\psi := |\psi\rangle$ is a spinor and $\phi^\dagger := \langle\phi|$ is a dual spinor such that $\langle\phi|\psi\rangle$ is a bilinear mapping into the complex numbers. We also define the Dirac adjoint $\bar{\psi} := \psi^\dagger \gamma^0 = \langle \psi|\gamma^0.$ Note that the notation $\bar{\psi}$ is reserved for the adjoint spinor, rather than complex conjugation, for the latter we utilise $*$ instead and write $\psi^*$.

\subsection{Helicity in Mathematics}
\subsubsection{Gamma matrices and the Clifford algebra $Cl_{1,3}$}

In this section, we will look at an alternative definition of helicity, which expresses the concept using the study of Clifford algebras~\cite{Lounesto,daRocha:2005ti}. This definition still makes sense with respect to the intuitive view of helicity as a projection of spin along the direction of motion, but the formalism used to describe this is radically different to that used in the previous section. Here, we will define it as a relation between a vector and a pseudo-vector type quantity which describes properties of Fermions in the context of the Dirac equation. These two quantities form part of a set of \emph{bilinear covariants} which are expressed as inner products of Dirac spinors and their adjoints. To establish the definition of helicity, it will be useful to recall the Dirac gamma matrices, and use these to generate the Clifford algebra $Cl_{1,3}$. We will then construct bilinear covariants, two of which we will used to define helicity in an alternative way.

Indeed, using the gamma matrices, we can generate a vector space isomorphic to the Clifford algebra $Cl_{1,3}$, which is useful because it can be thought of as the Clifford algebra of \emph{space-time} $\mathbb{R}^{1,3}$. Using elements of this Clifford algebra, combined with Dirac spinors, we can define objects which are used by physicists to construct Lagrangians to model particles, and furthermore can be used to classify spinors. 

Clifford algebras are a type of non-commutative algebras. They can be thought of as generalisations of complex numbers and quaternions to arbitrarily high dimensions. In most cases, a Clifford algebra is generated by a vector space and a quadratic form, or metric $g$ over that space, whereby basis elements $\textbf{e}_i$ and $\textbf{e}_j$ satisfy the relation: $ \textbf{e}_i\textbf{e}_j+ \textbf{e}_j\textbf{e}_i= g_{ij}\mathbbm{1}_n$, where $n$ is the dimension of the vector space and $i,j = 0,1,\ldots,n-1$. Here, juxtaposition of elements is known as the \emph{Clifford product}, we refer the reader to~\cite{Lounesto}.

The gamma matrices are one such set that represents the Clifford algebra $Cl_{1,3}$ where we also note that $(\gamma^0)^2 = \mathbbm{1}_4$ and $ (\gamma^\alpha )^2 = -\mathbbm{1}_4$ for $\alpha = 1,2,3$. This shows that the gamma matrices indeed form a basis of the Clifford algebra $Cl_{1,3}$. This limits the number and type of possible combinations for basis elements of the algebra. It is perhaps best to summarise the basis in table form:

\begin{table}[!ht] 
\centering
\begin{tabular}{l|l}
Type & Elements \\ 
\hline
Scalar & 1 (one element) \\
Vector & $\gamma^i$ (four elements) \\
Tensor & $\gamma^i\gamma^j$ (six elements)\\
Axial vector (Pseudo-vector) & $\gamma^i\gamma^j\gamma^k$ (four elements)\\
Pseudo-scalar & $\gamma^i\gamma^j\gamma^k\gamma^l \sim \gamma^5$ (one element)
\end{tabular}
\caption{Basis elements of the Clifford algebra $Cl_{1,3}$.}
\label{table1}
\end{table}

Note that $\gamma^5$ is the same as $ \gamma^0\gamma^1\gamma^2\gamma^3$ up to a scalar factor $i$. This is what is meant when we write $\gamma^i\gamma^j\gamma^k\gamma^l \sim \gamma^5$. Obviously, we have 16 basis elements, linear combinations of which form the Clifford algebra $Cl_{1,3}$. These basis elements can be put into five distinct spaces given by the left hand side of the column. The number of elements in each space is given in general by the numbers in Pascal's triangle.

We will now define the \emph{bilinear covariants}: If we take $A$ to be an element of the above list of 16 bases, and $|\psi\rangle$ to be a Dirac spinor in $\mathbb{C}^4$, then the quantity $\langle \psi | \gamma^0 A | \psi \rangle$ is called a bilinear covariant. This is because it is linear in both arguments, i.e.~$|\psi\rangle$ and $\langle\psi|$, and transforms correctly when undergoing coordinate or frame transformations. The motivation for constructing these objects is that physicists require Lorentz invariant quantities to model the mechanics of relativistic particles, and these objects are then ideal candidates for this analysis.
 
Looking at the table above, we see that the 16 basis elements of $Cl_{1,3}$ are categorised into 5 types, i.e.~they live in 5 distinct spaces. In exactly the same way, there are 5 different types of bilinear covariant we can possibly construct, and we categorise them as follows
\begin{enumerate}
\item $\Omega_1 = \bar{\psi}\psi$, \emph{scalar} bilinear covariant.
\item $J^a = \bar{\psi}\gamma^a\psi$, \emph{vector} bilinear covariant.
\item $S^{ab} = \bar{\psi}i\gamma^a\gamma^b\psi$, \emph{tensor} bilinear covariant.
\item $K_a = \bar{\psi}\gamma_5\gamma_a\psi$, \emph{axial vector} bilinear covariant.
\item $\Omega_2 = \bar{\psi}\gamma^5\psi$, \emph{pseudo-scalar} bilinear covariant.
\end{enumerate}
Note here that $ \gamma_a= \gamma^b \eta_{ab}$ is the covariant form of $\gamma^a$.

We will not go through the discussion of the significance of each bilinear covariant, but we will give brief explanations of their interpretation. $\Omega_1$ and $\Omega_2$ being scalar quantities, have only 1 degree of freedom, and so can be though of as describing the exchange of spin-0 bosons. $J^a$, being a vector, is usually interpreted as the particle's probability current, and thus contains information about the direction of the particle's motion. $S^{ab}$ is the particle's electromagnetic moment density. The axial vector $K_a$ is thought of as a form of spin projection. For details on these interpretations, we refer~\cite[p.~137-138]{Lounesto}.

\subsubsection{Helicity as a relation between bilinear covariants}

Recall our intuitive definition of helicity as a projection of spin along the direction of motion. If we base ourselves on the interpretations of the bilinear covariants above, it is clear that we can in some way use $J^a$ and $K_a$ to express helicity in a new way. We will now examine how this is done. We define the the following quantities: $\mathbf{K} = K_a \gamma^a$ and $\mathbf{J} = J^a \gamma_a$. This amounts to constructing vectors in the vector space with basis vectors $\gamma^a$ and $\gamma_a$, and components $K_a$ or $J^a$, respectively.

\begin{definition}{Helicity and bilinear covariants~\cite[p.~163]{Lounesto}:}
Let $\mathbf{J} = J^a \gamma_a$ and $\mathbf{K} = K_a \gamma^a$ be the vectors constructed from the vector and axial vector bilinear covariants, respectively. We define helicity $h$ to be the number such that $\mathbf{K} = h\mathbf{J}$.
\label{def2}
\end{definition}

Therefore we can say that the spin projection is parallel to the particle's probability current if $h=1$, and the spin projection is anti-parallel to the particle's probability current if $h=-1$. Now, we have a new definition of helicity, here in the context of Clifford algebras and bilinear covariants. Note that we assume the existence of this quantity and do not discuss this issue further. Although both these definitions are intuitively similar and relate to projecting spin along the direction of motion, it is not at all obvious whether both definitions are equivalent. Indeed, we have started from different parts of Physics and Mathematics to arrive at similar concepts. Although there are clear links, there are a number of hurdles which need to be overcome in order to see precisely how they are related. This is what we will do in the next section which contains our main result.

\section{Main result}

We will now prove our main result: 
\begin{align*}
  \text{Definition}~\ref{def2} \Rightarrow \text{Definition}~\ref{def1}.
\end{align*}

\subsection{Bilinear covariant equation in matrix form}

We start by finding the explicit form of $\mathbf{K} = h \mathbf{J}$. To do this, we firstly change to the Dirac notation where we denote
\begin{align} 
  |\psi\rangle = 
  \begin{pmatrix} 
    a \\ b \\ c \\ d 
  \end{pmatrix}, \qquad 
  \langle\psi| = 
  \begin{pmatrix} 
    a^* & b^* & c^* & d^* 
  \end{pmatrix},
\end{align}
where $a,b,c,d \in \mathbb{C}$. Let us begin with the definition on helicity in terms of bilinear covariants
\begin{align}
  \textbf{K} &= h \textbf{J}, \\
  K_a \gamma^a &= h J^b \gamma_b, \\
  \langle\psi| \gamma_0\gamma_5\gamma_a |\psi\rangle \gamma^a &= h \langle\psi| \gamma_0\gamma^b |\psi\rangle \gamma_b.
\end{align}
Expanding this expression in terms of Dirac matrices matrices, we arrive at\begin{footnote}{This was done by hand, and double checked using Mathematica. The code is available upon request.}\end{footnote}

\begin{multline}
  \begin{pmatrix} 
    0 & 0 & 2 a^* a - 2 d^* d & 2b^*a + 2d^*c \\
    0 & 0 & 2a^*b + 2 c^*d & 2 b^* b- 2 c^* c \\
    2 b^*b - 2 c^*c & -2 b^*a - 2 d^*c & 0 & 0 \\
    - 2 a^*b - 2 c^*d & 2 a^*a - 2 d^*d & 0 & 0 
  \end{pmatrix}\\
  =  h  
  \begin{pmatrix} 
    0 & 0 & 2 a^* a + 2 d^* d & 2b^*a - 2d^*c \\
    0 & 0 & 2a^*b - 2 c^*d & 2 b^*b + 2 c^* c \\
    2 b^*b + 2 c^*c & -2 b^*a + 2 d^*c & 0 & 0 \\
    - 2 a^*b + 2 c^*d & 2 a^*a + 2 d^*d & 0 & 0 
 \end{pmatrix}. 
\label{eqn_31}
\end{multline}

\subsection{Projection and charge conjugation operators} 

We found an explicit form of the equation $\textbf{K} = h \textbf{J}$. However, the result~(\ref{eqn_31}) is rather involved and we wish to make it more manageable. To do this, we define the following quantities
\begin{align} 
  |\phi_R\rangle = 
  \begin{pmatrix} 
    a \\ b 
  \end{pmatrix},\qquad
  |\phi_L\rangle = 
  \begin{pmatrix} 
    c \\ d 
  \end{pmatrix}.  
\end{align}
Therefore, using a slight abuse of notation we can write
\begin{align}
  |\psi\rangle = 
  \begin{pmatrix}
    |\phi_R \rangle  \\ |\phi_L \rangle 
  \end{pmatrix}.
\end{align}
The motivation for this rewriting is that the bilinear covariants in Definition~\ref{def2} involve 4-entry spinors whereas the eigenvalue equation of Definition~\ref{def1} is formulated using two 2-entry spinors. Next, we look at our explicit matrix for $\textbf{K}$ in equation~(\ref{eqn_31}). We would like to reduce this $4\times4$ matrix into a $2\times2$ matrix with $2\times2$ matrix entries. In order to achieve this, we examine first the top right hand 4 entries of $\textbf{K}$. Note that this splitting of terms in this matrix is of course arbitrary and is one of the subtle points why our main result works in one direction only. However, we believe that this is more of a technicality than a true obstacle in proving our result in the other direction. We write
\begin{align} 
  \begin{pmatrix} 
    2 a^* a - 2 d^* d & 2b^*a + 2d^*c \\
    2a^*b +  2c^*d &  2b^*b - 2c^* c 
  \end{pmatrix}  
  = 2 
  \begin{pmatrix}  
    a^* a  & b^*a  \\
    a^*b &  b^*b 
  \end{pmatrix} 
  + 2 
  \begin{pmatrix}  
    - d^* d &  d^*c \\
    c^*d & -  c^* c 
  \end{pmatrix}. 
\end{align} 
and notice that 
\begin{align} 
  \begin{pmatrix}  
    a^* a  & b^*a  \\
    a^*b &  b^*b 
  \end{pmatrix} 
  = 
  \begin{pmatrix}  
    a \\ b 
  \end{pmatrix}  
  \begin{pmatrix}  
    a^* &  b^* 
  \end{pmatrix} 
  = 
  |\phi_R\rangle\langle\phi_R|.
\end{align}
This is the exterior product of $|\phi_R\rangle $ with itself. In quantum mechanics and operator theory such an object is called a \emph{projection operator}. Indeed, notice that for normalised vectors $|\phi\rangle$ with $\langle\phi|\phi\rangle=1$, the operator $|\phi\rangle\langle\phi|$ is idempotent: $(|\phi\rangle\langle\phi|)^2=|\phi\rangle\langle\phi|\phi\rangle\langle\phi|=(|\phi\rangle)(\langle\phi|\phi\rangle)(\langle\phi|)=|\phi\rangle\langle\phi|$. Therefore, as claimed, this is a projection.

We would like a similar expression in terms of projection operators for the second $ 2\times2$ matrix (bottom left of~(\ref{eqn_31})), and indeed, we have 
\begin{align}  
  \begin{pmatrix} 
    - d^* d &  d^*c \\
    c^*d & -  c^* c 
  \end{pmatrix} 
  = -
  \begin{pmatrix}
    -d^*  \\
    c^* 
  \end{pmatrix} 
  \begin{pmatrix}
    -d & c 
  \end{pmatrix}.
\end{align}
In order to rewrite this expression in a very convenient form, we need to define the concept of charge conjugation, see~\cite[p.~168]{Lounesto}.

\begin{definition}{Charge conjugation for 4-spinors} 
We define the charge conjugation operator $\mathcal{C}: \mathbb{C}^4 \rightarrow \mathbb{C}^4$ acting on 4-spinors to be the mapping
\begin{align}
  |\psi\rangle \mapsto -i\gamma^2 |\psi \rangle^*.
\end{align}
\end{definition}

The charge conjugation operator inverts the charge of the spinor $|\psi\rangle$, or more physically, it exchanges a particle for its anti-particle. Using the 2-spinor notation, we can rewrite the action of this operator as
\begin{align} 
  \mathcal{C} |\psi \rangle = \mathcal{C} 
  \begin{pmatrix} 
    |\phi_R \rangle \\ 
    |\phi_L \rangle 
  \end{pmatrix} 
  = -i 
  \begin{pmatrix}
    \mathbb{O} & \sigma_2 \\ 
    -\sigma_2 & \mathbb{O} 
  \end{pmatrix}
  \begin{pmatrix} 
    |\phi_R \rangle^* \\ 
    |\phi_L \rangle^* 
  \end{pmatrix} 
  = 
  \begin{pmatrix} 
    -i\sigma_2 |\phi_R \rangle^* \\ 
    +i \sigma_2 | \phi_L \rangle^* 
  \end{pmatrix}.
\end{align}
This leads us to define a similar operator for 2-spinors.
\begin{definition}{Charge conjugation for 2-spinors}
We define the charge conjugation operator $C: \mathbb{C}^2 \rightarrow \mathbb{C}^2$ acting on 2-spinors to be the mapping 
\begin{align}
  |\phi\rangle \mapsto -i\sigma_2 |\phi \rangle ^*. 
\end{align}
\end{definition}

With this notation in place, we have
\begin{align} 
  \mathcal{C} |\psi\rangle = 
  \begin{pmatrix} 
    C|\phi_L \rangle \\ 
    -C|\phi_R \rangle 
  \end{pmatrix}.
\end{align}
This newly defined operator $C$ will be very important in the following, and we will use the shorthand $C|\phi\rangle = |\phi^C\rangle$.

\subsection{2-spinor representation}

Now, for some spinor $|\chi\rangle$ with components $\xi, \eta \in \mathbb{C}$, the charge conjugated spinor $|\chi^C\rangle $ satisfies
\begin{align} 
  |\chi^C\rangle = - \sigma_2 | \chi \rangle^* = -i 
  \begin{pmatrix} 
    0 & -i \\ i & 0 
  \end{pmatrix} 
  \begin{pmatrix} 
    \eta^* \\ \xi^* 
  \end{pmatrix} =
  \begin{pmatrix} 
    0 & -1 \\ 1 & 0 
  \end{pmatrix} 
  \begin{pmatrix} 
    \eta^* \\ \xi^* 
  \end{pmatrix} = 
  \begin{pmatrix} 
    -\xi^* \\ \eta^* 
  \end{pmatrix}, 
\end{align}
and we notice that 
\begin{align} 
  \begin{pmatrix}  
    - d^* d &  d^*c \\ c^*d & -  c^* c 
  \end{pmatrix} = (-1) 
  \begin{pmatrix}
    -d^* \\ c^* 
  \end{pmatrix} 
  \begin{pmatrix}  
    -d &  c 
  \end{pmatrix} 
  = - | \phi_L^C \rangle \langle \phi_L^C |.
\end{align}
In turn, we can thus rewrite the top right $2\times2$ matrix in $\textbf{K}$~(\ref{eqn_31}) as follows
\begin{multline} 
  \begin{pmatrix} 
    2 a^* a - 2 d^* d & 2b^*a + 2d^*c \\
    2a^*b +  2c^*d &  2b^*b - 2c^* c 
  \end{pmatrix}  
  = 2 
  \begin{pmatrix}  
    a^* a  & b^*a  \\
    a^*b &  b^*b 
  \end{pmatrix} 
  \\+ 2 
  \begin{pmatrix}  
    - d^* d &  d^*c \\
    c^*d & -  c^* c 
  \end{pmatrix} 
  = 2 | \phi_R \rangle \langle \phi_R | -2 | \phi_L^C \rangle \langle \phi^C_L|.
\end{multline}
We perform a similar substitution for the bottom left entries of $\textbf{K}$ and find
\begin{multline}
  \begin{pmatrix} 
    2 b^*b - 2 c^*c & -2 b^*a - 2 d^*c  \\
    - 2 a^*b - 2 c^*d & 2 a^*a - 2 d^*d 
  \end{pmatrix} 
  = 2 
  \begin{pmatrix} 
    b^*b & - b^*a \\ 
    - a^*b & a^*a 
  \end{pmatrix} 
  + 2 
  \begin{pmatrix} 
    -c^*c & - d^*c \\ 
    - c^*d & - d^*d 
  \end{pmatrix}
  \\= 2 
  \begin{pmatrix} 
    -b^* \\ a^* 
  \end{pmatrix} 
  \begin{pmatrix} 
    -b & a 
  \end{pmatrix} 
  -2 
  \begin{pmatrix} 
    c \\ d 
  \end{pmatrix} 
  \begin{pmatrix} 
    c^* & d^* 
  \end{pmatrix} 
  = 2 | \phi^C_R \rangle \langle \phi_R^ C| - 2 | \phi_L \rangle \langle \phi_L |.
\end{multline}
 
Therefore, we can now write $\textbf{K}$ concisely as a $2\times2$ matrix with $2\times2$ matrix entries
\begin{align} 
  \textbf{K} = 2 
  \begin{pmatrix} 
    \mathbb{O}  & | \phi_R \rangle \langle \phi_R | - | \phi_L^C \rangle \langle \phi^C_L|  \\ 
    | \phi^C_R \rangle \langle \phi_R^ C| -  | \phi_L \rangle \langle \phi_L |  & \mathbb{O} 
  \end{pmatrix}. 
  \label{Kmatrix}
\end{align}
This process can be repeated for the quantity $\textbf{J}$, see the right-hand side of~(\ref{eqn_31}). We find
\begin{align} 
  \textbf{J} = 2 
  \begin{pmatrix} 
    \mathbb{O}  & | \phi_R \rangle \langle  \phi_R | + | \phi_L^C \rangle \langle \phi_L^C | \\ 
    | \phi_L\rangle \langle \phi_L | + | \phi_R^C \rangle \langle \phi_R^C | &\mathbb{O}  
  \end{pmatrix}. 
\end{align}
Therefore, equation~(\ref{eqn_31}) is equivalent to
\begin{multline}
  \begin{pmatrix} 
    \mathbb{O}  & | \phi_R \rangle \langle \phi_R | - | \phi_L^C \rangle \langle \phi^C_L| \\ 
    | \phi^C_R \rangle \langle \phi_R^ C| - | \phi_L \rangle \langle \phi_L | & \mathbb{O}  
  \end{pmatrix}
  \\ = h  
  \begin{pmatrix} 
    \mathbb{O}  & | \phi_R \rangle \langle  \phi_R | + | \phi_L^C \rangle \langle \phi_L ^C| \\ 
    | \phi_L \rangle \langle \phi_L| + | \phi_R^C \rangle \langle \phi_R^C | & \mathbb{O}  
  \end{pmatrix}.
  \label{eq_3.18}
\end{multline}

Before proceeding, it is worth making a short comment about the orthogonality of the 2-entry spinors we are considering. Indeed, it is interesting to note that 
\begin{align} 
  \langle \phi | \phi^C \rangle = 
  \begin{pmatrix} 
    \eta* & \xi^* 
  \end{pmatrix} 
  \begin{pmatrix} 
    -\xi^* \\ \eta^* 
  \end{pmatrix} 
  = -\eta^* \xi^* + \xi^* \eta^* = 0. 
\end{align}
This means that a state and its conjugate are orthogonal by construction. 

Therefore, considering the $h=1$ case in~(\ref{eq_3.18}) means to consider a purely right-handed state, i.e.~setting $c = d = 0$ in the momentum space representation of $|\phi_L \rangle$, which in turn means that $|\phi_L \rangle = |\phi_R^C \rangle = 0 $. Conversely, taking $h=-1$ means considering a totally left-handed state, i.e.~we set $a=b=0$, which implies $|\phi_R\rangle = | \phi_L^C \rangle =0$. We will consider each of these two cases separately.

\subsection{Positive helicity $h=1$}

As stated above, we begin with the case $h=1$ which is equivalent to setting $c=d=0$. This means that $|\phi_L\rangle = |\phi_L^C\rangle = 0$. Therefore, equation~(\ref{eq_3.18}) reduces to a simple identity. It is at this stage that we will invoke the momentum space representation. This will allow us to rewrite the projection operators in terms of the quantum mechanical helicity operator which is at the heart of Definition~\ref{def1}. Recall the explicit representation of $|\phi_R\rangle$ in~(\ref{psiR}). We can use this to calculate the explicit representation of $|\phi_R^C\rangle$ which is given by
\begin{align}
  |\phi_R^C\rangle = -i\sigma_2 |\phi_R\rangle^* = 
  \begin{pmatrix} 
    -\sin (\theta/2) e^{-i\varphi /2} \\  
    \cos (\theta /2)  e^{i \varphi /2 } 
  \end{pmatrix}.
\end{align}
Next, we wish to compute an explicit representation of the projectors $|\phi_R\rangle \langle\phi_R|$ and $|\phi_R^C\rangle \langle\phi_R^C|$ in the momentum space representation. We begin with $|\phi_R\rangle \langle \phi_R|$ which becomes
\begin{align}
  |\phi_R\rangle \langle\phi_R| &= 
  \begin{pmatrix} 
    \cos (\theta/2) e^{-i\varphi /2} \\
    \sin (\theta /2)  e^{i \varphi /2 } 
  \end{pmatrix} 
  \begin{pmatrix} 
    \cos (\theta/2) e^{i\varphi /2}  & \sin (\theta /2)  e^{-i \varphi /2 } 
  \end{pmatrix} 
  \nonumber\\
  &=   
  \begin{pmatrix} 
    \cos^2(\theta/2) & \cos(\theta/2)\sin(\theta/2) e^{-i\varphi} \\
    \cos(\theta/2)\sin(\theta/2) e^{i\varphi} & \sin^2(\theta/2) 
  \end{pmatrix}.
\end{align}
We now apply double angular formulae and obtain
\begin{align}
  |\phi_R\rangle \langle\phi_R| &= \frac{1}{2} 
  \begin{pmatrix} 
    (1+\cos{\theta}) & \sin{\theta} e^{-i\varphi} \\ 
    \sin{\theta} e^{i\varphi} & (1 - \cos{\theta}) 
  \end{pmatrix} 
  \nonumber \\
  &= \frac{1}{2} 
  \begin{pmatrix} 
    1 & 0 \\ 0 & 1 
  \end{pmatrix} 
  + \frac{1}{2} 
  \begin{pmatrix} 
    \cos(\theta) & \sin(\theta) e^{-i\varphi} \\ 
    \sin(\theta) e^{i\varphi} & - \cos(\theta) 
  \end{pmatrix}.
\end{align}
Comparing this expression with the helicity operator~(\ref{helicity}), we write
\begin{align} 
  |\phi_R\rangle \langle\phi_R | = 
  \frac{1}{2} (\mathbbm{1} + (\vec{\sigma} \cdot \hat{p})). 
  \label{first}
\end{align}
Likewise, for $|\phi_R^C\rangle \langle\phi_R^C|$ we find 
\begin{align}
  |\phi_R^C\rangle \langle\phi_R^C| = 
  \frac{1}{2} ( \mathbbm{1} - (\vec{\sigma} \cdot \hat{p})).
  \label{second}
\end{align}

In order to make the link with helicity, it suffices to multiply~(\ref{first}) by $|\phi_R\rangle$ from the right, and~(\ref{second}) by $|\phi_R^C\rangle$, also from the right. Since we assume our states to be normalised, this yields
\begin{align} 
  |\phi_R\rangle &= \frac{1}{2} (\mathbbm{1} + (\vec{\sigma} \cdot \hat{p}))|\phi_R\rangle,
  \label{firstb}\\
  |\phi_R^C\rangle &= \frac{1}{2} ( \mathbbm{1} - (\vec{\sigma} \cdot \hat{p})) |\phi_R^C\rangle.
  \label{secondb}
\end{align}
A simple step of algebra results in the two equations
\begin{align} 
  (\vec{\sigma} \cdot \hat{p} )| \phi_R \rangle &= +1 | \phi_R \rangle,
  \label{firstc}\\
  (\vec{\sigma} \cdot \hat{p} )| \phi_R^C \rangle &= -1 | \phi_R^C \rangle.
  \label{secondc}
\end{align}
This means that we recovered the correct definition for the quantum mechanical helicity operator acting on the right-handed states! We additionally obtain an extra equation which explains how the conjugate states behave (we will interpret this later).

\subsection{Negative helicity $h=-1$}

The calculation of the previous subsection can easily be repeated for the $h=-1$ case. We find
\begin{align}
  |\phi_L^C \rangle \langle \phi_L^C| &= \frac{1}{2} ( \mathbbm{1} + (\vec{\sigma} \cdot \hat{p})),
  \label{h1}\\ 
  |\phi_L \rangle \langle \phi_L | &= \frac{1}{2} ( \mathbbm{1} - ( \vec{\sigma} \cdot \hat{p}) ). 
  \label{h2}
\end{align}
If we, as before, multiply these two equations by $|\phi_L^C \rangle$ and $|\phi_L \rangle$, respectively, we arrive at
\begin{align} 
  (\vec{\sigma} \cdot \hat{p} )| \phi_L \rangle &= -1 | \phi_L \rangle,
  \label{firstd}\\
  (\vec{\sigma} \cdot \hat{p} )| \phi_L^C \rangle &= +1 | \phi_L^C \rangle.
  \label{secondd}
\end{align}
Indeed, this gives us the other part of the helicity operator definition which we were looking for.

\subsection{Summary}

Collecting Eqs.~(\ref{firstc})--(\ref{secondc}) and~(\ref{firstd})--(\ref{secondd}) gives
\begin{align}
  (\vec{\sigma} \cdot \hat{p} )| \phi_R \rangle &= +1 | \phi_R \rangle,
  \label{f1}\\
  (\vec{\sigma} \cdot \hat{p} )| \phi_L \rangle &= -1 | \phi_L \rangle,
  \label{f2}\\
  (\vec{\sigma} \cdot \hat{p} )| \phi_R^C \rangle &= -1 | \phi_R^C \rangle,
  \label{f3}\\
  (\vec{\sigma} \cdot \hat{p} )| \phi_L^C \rangle &= +1 | \phi_L^C \rangle.
  \label{f4}
\end{align}
The first two of these equations~(\ref{f1}) and~(\ref{f2}) are exactly the relations which define helicity as an eigenvalue equation, see Definition~\ref{def1}. The second pair of equations~(\ref{f3}) and~(\ref{f4}) needs some interpretation. Note that $(\vec{\sigma}\cdot\hat{p})|\phi_{R/L}\rangle = \pm|\phi_{R/L}\rangle$ is equivalent to $(\vec{\sigma}\cdot\hat{p})|\phi_{R/L}^C\rangle = \mp|\phi_{R/L}^C\rangle$ under the action of the charge conjugation operator $\mathcal{C}$. This means that the second pair of equations are simply the equivalent definition of helicity for anti-particles. Thus, we have shown that the definition of helicity in terms of bilinear covariants $\textbf{K} = h \textbf{J}$ implies the definition of helicity in quantum mechanics $(\vec{\sigma}\cdot\hat{p})|\phi_{R/L}\rangle = \pm|\phi_{R/L}\rangle$ and the equivalent relation for anti-particles, which is what we had set out to establish. Thus our main result is proved.

\section{Applications to the study of Graphene}

The discovery and synthesis of Graphene is a recent development in solid state physics. It has attracted much attention from the scientific community, for instance due to its superconductive nature. The Physics Nobel Prize of 2010 was awarded to Geim and Novoselov for the synthesis of Graphene~\cite{Nobel}, and much research is still focusing on it today. Graphene is a one-layer honeycomb-shaped lattice of carbon molecules. The properties of graphitic structures have been heavily studied in the past, mainly in the context of fullerenes and carbon nanotubes, and the study of Graphene has benefited much from this, as one can view these particular structures as different arrangements of Graphene. Perhaps the most notable property of Graphene, and the one which undoubtedly has the most exciting direct applications, is its response to electrons and current flow. Indeed, Graphene acts as a superconductor, and so the idea of modeling current flow through Graphene using the Dirac spinor formalism has become relatively widespread in this area of physics~\cite{Abergel:2010kw}. Remarkably, it is the helicity operator that plays a pivotal role in the Hamiltonian for 1-layer Graphene.

\subsection{Graphene Hamiltonian in terms of helicity}

As expressed in the recent review~\cite[p.~270]{Abergel:2010kw}, in a tight-binding model of monolayer Graphene the equations governing the motion of charges can be viewed as those of relativistic Dirac Fermions on a plane (here, that of the Graphene). Using this model, one can formulate the Hamiltonian for particles in the vicinity of a vertex $V$ in one of the hexagonal lattices of Graphene
\begin{align} 
  \hat{\mathcal{H}}_V ( \vec{k} ) = \hbar v_F 
  \begin{pmatrix} 
    0 & k_x - i k_y \\ k_x + i k_y & 0 
  \end{pmatrix}.
  \label{ham1}
\end{align}
Here, $\vec{k} $ is the particle's wavevector along its direction of motion with components $k_x,k_y$ in the Graphene plane and $\hbar$ is Planck's reduced constant. The speed of the Fermions is denoted by $v_F$. Since one is working in the plane, rather than in space, the Pauli spin vector can be redefined in this context, omitting the $z$-direction. Therefore 
\begin{align} 
  \vec{k} = 
  \begin{pmatrix} 
    k_x \\ k_y 
  \end{pmatrix},
  \qquad \vec{\sigma} = 
  \begin{pmatrix} 
    \sigma_1 \\ \sigma_2 
  \end{pmatrix}. 
\end{align}
Since $\vec{k}/|\vec{k}|$ is a unit vector along the direction of motion of the charge, we can rewrite the above Hamiltonian~(\ref{ham1}) as
\begin{align} 
  \hat{\mathcal{H}}_V ( \vec{k} ) = 
  \hbar v_F (\vec{\sigma} \cdot \vec{k} ) = 
  \frac{\hbar v_F}{|\vec{k}|} (\vec{\sigma} \cdot \vec{k}).
  \label{ham2}
\end{align}
Thus, the Hamiltonian for Fermions in the vicinity of $V$ is proportional to the two-dimensional helicity operator. This remarkable occurrence of the helicity operator allows us to apply some of the formalism we used in the previous Sections to re-write the Hamiltonian~(\ref{ham2}) in a new way.

\subsection{Rewriting the Graphene Hamiltonian}

Recalling Eqs.~(\ref{h1}) and~(\ref{h2}) we notice that
\begin{align} 
  | \phi_L^C \rangle \langle \phi_L^C | - | \phi_L \rangle \langle \phi_L | = 
  \frac{1}{2} (\mathbbm{1} + (\vec{\sigma} \cdot \hat{p})) - 
  \frac{1}{2} (\mathbbm{1} - (\vec{\sigma} \cdot \hat{p})) = 
  (\vec{\sigma} \cdot \hat{p}) .
\end{align}
Therefore, the Hamiltonian~(\ref{ham1}) can be written as 
\begin{align}
  \hat{\mathcal{H}}_V (\vec{k}) = 
  \frac{\hbar v_F}{|\vec{k}|} 
  \bigl(| \phi_L^C \rangle \langle \phi_L^C | - | \phi_L \rangle \langle \phi_L| \bigr). 
\end{align} 
This is a neat preliminary result, as it allows us to write the Hamiltonian in terms of the possible eigenstates of a Fermion travelling trough Graphene near the vertex $V$. Note that we could equally have chosen the other projectors $|\phi_R \rangle \langle \phi_R|$ and $ | \phi_R^C \rangle \langle \phi_R^C|$, safe in the knowledge that both projectors reside in the same space and that arbitrary phase factors between left-handed and right-handed states do not have any physical effects. 

Now, we would like to relate this to the operator $\textbf{K}$ which was at heart of our work. Since we chose to write~(\ref{ham2}) in terms of the left projector and its conjugate, we inspect the form of $\textbf{K}$ when considering left-handed states, this means setting $|\phi_R\rangle = 0$. Eq.~(\ref{Kmatrix}) yields
\begin{align}
  \textbf{K}_{_{(|\phi_R\rangle =0)}} = 2 
  \begin{pmatrix} 
    \mathbb{O} & -|\phi_L^C \rangle \langle \phi_L^C| \\ 
    - |\phi_L \rangle \langle \phi_L | & \mathbb{O} 
  \end{pmatrix}.
\end{align}
The matrix $\gamma^5 = i\gamma^0\gamma^1\gamma^2\gamma^3$ in the Weyl basis is given by
\begin{align} 
  \gamma^5 = 
  \begin{pmatrix} 
    -\mathbbm{1} & \mathbb{O} \\ 
    \mathbb{O} & \mathbbm{1} 
  \end{pmatrix}, 
\end{align}
which can be used for the following rewriting
\begin{align} 
  \gamma^5\textbf{K}_{_{(|\phi_R\rangle =0)}} \gamma^0  = 2 
  \begin{pmatrix} 
    |\phi_L^C \rangle \langle \phi_L^C | & \mathbb{O} \\ 
    \mathbb{O} & - | \phi_L \rangle \langle \phi_L| 
  \end{pmatrix}.
\end{align}
Taking the trace of this last equation results in 
\begin{align} 
  \hat{\mathcal{H}}_V ( \vec{k} ) = 
  \frac{ \hbar v_F}{2|\vec{k}|}\, \mathrm{tr} \bigl(\gamma^5\textbf{K}_{_{(|\phi_R\rangle =0)}}\gamma^0\bigr),
\end{align}
which is a very surprising and unexpected result as it allows us to formulate the Graphene Hamiltonian in terms of objects used in the classification of Clifford algebras. Note that we can safely use the 3 dimensional version of $\textbf{K}$ since the helicity operator we started off with, i.e.~$(\vec{\sigma} \cdot \hat{k})$ is still the same as the 3 dimensional operator, only with the $z$-component of $\vec{k}$ set to zero. This is a fascinating preliminary result, we are happy to admit that we do not understand yet the full extent of these findings but think that research along these lines might be fruitful. 

\section{Conclusions}

We have shown that the mathematical definition of helicity based on objects used to classify Clifford algebras implies the physical definition of helicity. The more mathematically-oriented definition is slightly more general than the physical one since it naturally encompasses relations for both particles and anti-particles. It would be interesting to perform the same calculation in reverse direction. Conceptually, there should be very little stopping us from showing equivalence between the two definitions. 

An unexpected side result of our work was concerned with Graphene. We were able to reformulate its Hamiltonian in an intriguing way which might open new avenues of further research. It seems remarkable that such a result should have cropped up in a project dealing chiefly the study of Clifford algebras and helicity. Then again, helicity has been an interesting topic for many decades. Its ubiquity and importance in so many fields of research was the reason that disjoint definitions of it appeared in the literature, our main task was to show that these definitions indeed describe the same properties of nature.

\section*{Acknowledgements}
We would like to thank Yuri Obukhov for helpful discussions.

\end{document}